\newif\ifws
\ifws

\documentclass[entropy,article,submit,pdftex,oneauthor]{Definitions/mdpi}
\firstpage{1}
\pubvolume{1}
\issuenum{1}
\articlenumber{0}
\pubyear{2025}
\copyrightyear{2025}
\datereceived{ }
\daterevised{ } % Comment out if no revised date
\dateaccepted{ }
\datepublished{ }
\hreflink{https://doi.org/} % If needed use \linebreak
\Title{Chromatic Quantum Contextuality}

\TitleCitation{Chromatic Quantum Contextuality}

\Author{Karl Svozil $^{1,\dagger,\ddagger}$\orcidA{}}

\AuthorNames{Karl Svozil}

\isAPAStyle{%
       \AuthorCitation{Svozil, K.}
         }{%
        \isChicagoStyle{%
        \AuthorCitation{Svozil,Karl}
        }{
        \AuthorCitation{Svozil, K.}
        }
}

\address{%
$^{1}$ \quad Institute for Theoretical Physics,
TU Wien,
Wiedner Hauptstrasse 8-10/136,
1040 Vienna,  Austria; karl.svozil@tuwien.ac.at}

\corres{Correspondence: karl.svozil@tuwien.ac.at}

\abstract{
Chromatic quantum contextuality is a criterion of quantum nonclassicality based on (hyper)graph coloring constraints. If a quantum hypergraph requires more colors than the number of outcomes per maximal observable (context), it lacks a classical realization with $n$-uniform outcomes per context. Consequently, it cannot represent a `completable' non-contextual set of coexisting $n$-ary outcomes per maximal observable. This result serves as a chromatic analogue of the Kochen-Specker theorem. We present an explicit example of a four-colorable quantum logic in dimension three. Furthermore, chromatic contextuality suggests a novel restriction on classical truth values, thereby excluding two-valued measures that cannot be extended to $n$-ary colorings. Using this framework, we establish new bounds for the house, pentagon, and pentagram hypergraphs, refining previous constraints.
}


\usepackage{tikz}
\usetikzlibrary{decorations.markings}
\usetikzlibrary{calc}

\begin{document}

%%%%%%%%%%%%%%%%%%%%%%%%%%%%%%%%%%%%%%%%%%

\else

\documentclass[%
 %reprint,
 superscriptaddress,
 %groupedaddress,
 %unsortedaddress,
 %runinaddress,
 %frontmatterverbose,
  reprint,
 %onecolumn,
 showpacs,
 showkeys,
 %preprintnumbers,
 nofootinbib,
 %nobibnotes,
 %bibnotes,
  amsmath,amssymb,
 % aps,
 % prl,
 pra,
 %prb,
 % rmp,
 %prstab,
 %prstper,
  longbibliography,
  floatfix,
  %lengthcheck,%
 ]{revtex4-2}

\usepackage[normalem]{ulem}

\usepackage{adjustbox}

\usepackage{hyperref}
\usepackage{amsmath}
\usepackage{amssymb}
\usepackage{amsthm}
\usepackage{bm} %bold math
\usepackage{graphicx}

\RequirePackage{times}
\RequirePackage{mathptm}

\usepackage{url}
\usepackage[x11names]{xcolor}
\usepackage{eepic}
\usepackage{tikz}
\usetikzlibrary{decorations.markings}
\usetikzlibrary{calc}
\usepackage {pgfplots}
\pgfplotsset {compat=1.8}
\usepackage{epstopdf}
\usepackage[normalem]{ulem}
\sloppy

\newtheorem{theorem}{Theorem}
\newtheorem{comment}{Comment}
\newtheorem{proposition}{Proposition}
\newtheorem{corollary}{Corollary}
\newtheorem{fact}{Fact}
\newtheorem{lemma}{Lemma}
\theoremstyle{definition}
\newtheorem{definition}{Definition}

\newcommand{\seq}[1]{\mathbf{#1}}
\newcommand{\floor}[1]{\left\lfloor #1 \right\rfloor}
\newcommand{\ceil}[1]{\left\lceil #1 \right\rceil}
\newcommand{\abs}[1]{\left\lvert#1\right\rvert}
\newcommand{\rest}[2]{#1\!\!\restriction_{#2}}
\newcommand{\reste}[2]{#1\restriction_{#2}}
\newcommand{\N}{\mathbb{N}}%      \N   == \mathbb{N}
\newcommand{\Z}{\mathbb{Z}}%      \Z   == \mathbb{Z}
\newcommand{\Q}{\mathbb{Q}}%      \Q   == \mathbb{Q}
\newcommand{\R}{\mathbb{R}}%      \R   == \mathbb{R}
\newcommand{\C}{\mathbb{C}}
\newcommand{\alphabet}{\{0,1\}}
\newcommand{\B}{B^*}%        \X  == \Sigma^*
\newcommand{\BI}{B^\omega}%        \XI  == \Sigma^\infty
\newcommand{\x}{\mathbf{x}}
\newcommand{\dom}{\text{dom}}
\newcommand{\cl}{\text{cl}}
\newcommand{\dd}{\mathrm{d}}

\newcommand{\bra}[1]{\left< #1 \right|}
\newcommand{\ket}[1]{\left| #1 \right>}

\newcommand{\iprod}[2]{\langle #1 | #2 \rangle}
\newcommand{\mprod}[3]{\langle #1 | #2 | #3 \rangle}
\newcommand{\oprod}[2]{| #1 \rangle\langle #2 |}

\newcommand{%!
\new}[1]{{\color{magenta}#1}}

\begin{document}

%\title{Infinity Does It: Macroscopic Irreversibility From Microscopic Reversibility by Infinite Means}
\title{(FAPP) Infinity Does Macroscopic Irreversibility From Microscopic Reversibility}
%Quantum random number generator by value indefiniteness of one observable and its remainders within a context}

\author{Karl Svozil}
\email{karl.svozil@tuwien.ac.at}
\homepage{http://tph.tuwien.ac.at/~svozil}

\affiliation{Institute for Theoretical Physics,
TU Wien,
Wiedner Hauptstra{\ss}e 8-10/136,
1040 Vienna,  Austria}

\date{\today}

\begin{abstract}
Infinity is central to deriving macroscopic irreversibility from reversible microscopic laws across mathematics, theoretical computer science and physics.
In analysis, infinite processes---such as Dedekind cuts and Cauchy sequences---construct real numbers as equivalence classes of rational approximations, bridging discrete rationals to the continuous real line.
In quantum mechanics, infinite tensor products model nested measurements, where sectorization partitions the Hilbert space into equivalence classes, reconciling unitary evolution with wavefunction collapse.
In statistical mechanics, macrostates emerge as equivalence classes of microstates sharing identical macroscopic properties, providing the statistical basis for thermodynamic irreversibility despite reversible dynamics.
Equivalence relations formalize For-All-Practical-Purposes (FAPP) indistinguishability, reflecting operational limits on precision and observation.
Together, these examples reveal a unified framework where infinity and equivalence underpin emergent macroscopic behavior from microscopic reversibility.
\end{abstract}

%\pacs{03.65.Aa, 03.65.Ta, 03.65.Ud, 03.67.-a}

\maketitle

\fi

% Please write a section of an article, in LaTeX markup language, about the various methods to go from the rational numbers to the reals by infinite means, emphasizing the requirement of infinity. Thereby, also mention Cantor diagonalization as a way to construct---through an infinite process--an irrational number from the rationals by infinite means.

\section{From Rationals to Reals: The Role of Infinity}

The construction of the real numbers from the rational numbers is a fundamental topic in mathematical analysis, highlighting the necessity of infinite processes. The rational numbers, denoted by \(\mathbb{Q}\),
are countable and dense in the real numbers, but they are incomplete. This incompleteness arises because there exist `gaps' in \(\mathbb{Q}\)
that correspond to irrational numbers. To fill these gaps and construct the real numbers, \(\mathbb{R}\), mathematicians employ infinite methods, such as continued fractions.
Two prominent approaches are \textit{Dedekind cuts} and \textit{Cauchy sequences}, both of which rely on the concept of infinity.
The discussion will explore methods employing infinite means that transcend from rational to irrational numbers, then progress through Specker sequences to uncomputable numbers, and finally examine Omega sequences leading to algorithmically incompressible random reals.
At this point, concerns about the physical operationality of these infinite means will be set aside, with the issue revisited later in the discussion.

\subsection{Dedekind Cuts}
A Dedekind cut partitions the rational numbers into two non-empty sets \(A\) and \(B\) such that every element of \(A\) is less than every element of \(B\), and \(A\) contains no greatest element. The cut represents a real number, which may be rational or irrational. For example, the cut corresponding to \(\sqrt{2}\) is defined by:
\[
A = \{ x \in \mathbb{Q} \mid x^2 < 2 \}, \quad B = \{ x \in \mathbb{Q} \mid x^2 > 2 \}.
\]
This construction inherently involves an infinite sets \(A\) and \(B\).
The completeness of the real continuum is embodied by the property that every such cut corresponds to a unique real number, effectively filling the `irrational gaps between' rationals.
This construction vividly illustrates that the limit of a
sequence---often an irrational number---can be captured only through an infinite process corresponding to the infinite sets \(A\) and \(B\).

Similarly, surreal numbers, introduced by Conway~\cite{conway-ONAG} and explored in a mathematical dialogue by Knuth~\cite{knuth-surreal-numbers},
are constructed recursively as equivalence classes of pairs of sets of surreal numbers, subject to the condition that every element of the first set is less than every element of the second set.
The construction begins with the empty set. At each stage, new numbers are defined as $\{ L \mid R \}$, where $L$ and $R$ are sets of previously constructed numbers, provided that every member of $L$ is less than every member of $R$.
This Dedekind cut-like procedure, iterated transfinitely and allowing $L$ and $R$ to be infinite, produces not only all standard real numbers but also a vast continuum of infinite and infinitesimal numbers.
Thus, from the initial void---the empty set $\{\, \mid \,\}$ identified with the number $0$---this infinite process generates a comprehensive universe of numbers, truly \textit{ex nihilo omnia} (everything out of nothing).

\subsection{Cauchy Sequences}
Another method to construct the real numbers is through Cauchy sequences of rational numbers. A Cauchy sequence \((x_n)_{n=1}^{\infty}\) is a sequence whose elements become arbitrarily close to each other as the sequence progresses. Formally, for every \(\epsilon > 0\), there exists an integer \(N\) such that for all \(m, n \geq N\), \(|x_m - x_n| < \epsilon\). The real numbers are then defined as equivalence classes of Cauchy sequences, where two sequences are equivalent if their difference converges to zero. This process also relies on infinity, as the convergence of the sequence is an infinite phenomenon.

\subsection{Infinite Decimal Expansions}
A more familiar representation is that of \emph{infinite decimal expansions}. Any real number can be expressed as an infinite sequence of digits,
\(
x_0.x_1x_2x_3\ldots
\),
which in turn can be viewed as an infinite sum. This representation not only emphasizes the necessity of an infinite process but also shows how numbers that cannot be finitely represented (such as irrational numbers) naturally arise from the completion of an endless procedure.

\subsection{Cantor's Diagonalization and Irrational Numbers}
Cantor's diagonalization argument is a powerful tool that demonstrates the uncountability of the real numbers and provides a method to construct irrational numbers from rationals through an infinite process.
Consider an enumeration of all rational numbers in the interval \([0, 1]\), say \(r_1, r_2, r_3, \ldots\).
Each rational number \(r_i\) can be expressed as an infinite decimal expansion.
By constructing a new number \(x\) whose \(n\)-th decimal digit differs from the \(n\)-th decimal digit of \(r_n\),
we ensure that \(x\) is distinct from every rational number in the list.
For instance, if the \(n\)-th digit of \(r_n\) is \(d_n\), define the \(n\)-th digit of \(x\) as \(d_n + 1 \mod 10\).
The resulting number \(x\) is irrational, as it cannot correspond to any rational number in the enumeration.
This construction explicitly relies on an infinite process that `constructs' an irrational number~\cite{Yanofsky-BSL:9051621,bridgman}.

\subsection{No Continua Without Infinite Means}

The transition from the rational numbers to the real numbers---whether through Dedekind cuts, Cauchy sequences, or infinite decimal expansions---as well as Cantor's diagonalization argument necessitates the use of infinite means.
These Zeno-type
constructions underscore the indispensable role of infinity in bridging the gap between the countable realm of \(\mathbb{Q}\) and the uncountable continuum of \(\mathbb{R}\):
No finite procedure that starts with a finite set of rational numbers and uses only a finite number of operations can produce an irrational number.

Whether the infinities inherently present in (classical) continua can be put to any operational physical use remains an open question~\cite{svozil-set}.
Suffice it to say that the assumption of continua, as well as the selection of one of their elements via the axiom of choice, is a key ingredient in the apparent oxymoron that is the widely used term deterministic chaos.

Noson Yanofsky has noted that the procedural approach used here to generate the continuum and other mathematical entities, such as irrational or uncomputable numbers, including through methods like diagonalization, could be criticized.
The criticism stems from the view that tools like Dedekind cuts and Cauchy sequences describe or represent numbers rather than actually constructing them~\cite{Yanofsky2025-02-06-pc}].
However, while this raises a relevant metamathematical concern, it ultimately hinges on a matter of philosophical perspective.

%Write another subsection discussing Specker sequences that are sequences of computable numbers that have their limit as an uncomputable number, and mention Chaitin's Omega as one example of a Specker sequence. Threby, please emphasize the role of infinity

\subsection{Specker Sequences and the Role of Infinity}

Just as infinity plays an indispensable role in the transition from rational to irrational numbers and in the conceptualization of mathematical continua,
\textit{Specker sequences} provide a profound illustration of how infinity can lead us from the computable to the uncomputable,  thereby selecting a subset of irrationals by tightening criteria.
Almost all reals are of this type.

A Specker sequence is a computable, monotonically increasing, bounded sequence of rational numbers whose limit is an uncomputable real number~\cite{specker49,kreisel}.
Formally, a sequence $(a_n)_{n=1}^{\infty}$ is a Specker sequence if:
\begin{enumerate}
    \item Each $a_n$ is a computable rational number
    \item The sequence is strictly increasing: $a_n < a_{n+1}$ for all $n$
    \item The sequence is bounded above: there exists $L \in \mathbb{Q}$ such that $a_n < L$ for all $n$
    \item The limit $\lim_{n \to \infty} a_n$ is not a computable number
\end{enumerate}

The existence of such sequences demonstrates that the infinite completion of even well-behaved, computable objects can yield entities beyond algorithmic reach.
The essence of Specker's construction is to  encode an undecidable property into the convergence behavior of the sequence. Although each term \(a_n\) is produced by a finite, effective algorithm, the process of converging to \(L\) is intrinsically infinite---any
attempt to specify a convergence criterion would require solving a problem that is uncomputable---indeed, to quote an early, informal intuition by Paul Ehrenfest, such a convergence criterion ``grows beyond any specifiable size''~\cite{Ehrenfest_09c}.
 In this way, the limit \(L\) becomes an uncomputable real number even though it is the limit of a computable (recursive) sequence.

\subsection{Chaitin's Omega as the Ultimate Specker Sequence}

Perhaps the most profound example of a limit of a Specker sequence is Chaitin's Omega ($\Omega$), often called the `halting probability'~\cite{ch:75}.
This number represents the probability that a randomly constructed self-delimiting program will halt when run on a universal Turing machine.

Chaitin's $\Omega$ can be expressed as:

$$\Omega = \sum_{p \text{ halts}} 2^{-|p|}$$

where the sum is taken over all self-delimiting programs $p$ that halt, and $|p|$ denotes the length of program $p$ in bits.

$\Omega$ can be approximated by
rational numbers:

$$\Omega_n = \sum_{\substack{p \text{ halts within} \\ n \text{ steps}}} 2^{-|p|}$$

The sequence $(\Omega_n)_{n=1}^{\infty}$ is a Specker sequence---each $\Omega_n$ is computable for `small' $n$~\cite{calude-dinneen06}, the sequence is monotonically increasing, bounded above by 1, yet its limit $\Omega$ is uncomputable.
The uncomputable nature of $\Omega$ stems from the fact that knowledge of its binary expansion would allow us to solve the Halting Problem, which is provable impossible.
Each additional bit of precision in $\Omega$ encodes the solution to increasingly complex instances of the Halting Problem.
There does not exist any computable convergence criterion: just as for computing the $n$th bit of a Busy Beaver function, the time to compute those instances $\Omega_n$ outgrows any computable function of $n$~\cite{chaitin-bb}.

All of the above reveals a fundamental qualitative shift at infinity---one that goes beyond mere quantitative change:
Infinity generates fundamentally new mathematical objects. The rational numbers, all of which are computable, give rise through infinite processes to real numbers that no algorithm can fully capture.

The transition from the finite to the infinite marks a profound divide between what is algorithmically accessible and what remains beyond reach.
While we can approximate \(\Omega\) arbitrarily closely using computable methods, we can never compute it exactly.
Specker sequences thus demonstrate that infinity is not merely a convenient mathematical abstraction but a necessary concept that marks the boundary between the computable and the uncomputable,
between what can be algorithmically constructed and what can only be defined through infinite convergence.
Indeed, despite random reals~\cite{martin-lof} constituting almost all irrational numbers, locating specific instances through computational, finite, or physically operational means remains provably impossible~\cite{calude:02}.

Specker sequences and Chaitin's Omega are not defined using equivalence classes in their original formulations.
But they are related to equivalence classes in how they can be introduced: They are indirectly tied through the Cauchy sequence construction of real numbers, where their limits are equivalence classes.

%%%%%%%%%%%%%%%%%%%%%%%%%%%%%%%%%%%%%%%%%%%%%%%%%%%%%%%%%%%%%%%%%%%%%%%%%%%%%%%%%%%%%%%%%%%%%%%%%%%%%%%%
%%%%%%%%%%%%%%%%%%%%%%%%%%%%%%%%%%%%%%%%%%%%%%%%%%%%%%%%%%%%%%%%%%%%%%%%%%%%%%%%%%%%%%%%%%%%%%%%%%%%%%%%
%%%%%%%%%%%%%%%%%%%%%%%%%%%%%%%%%%%%%%%%%%%%%%%%%%%%%%%%%%%%%%%%%%%%%%%%%%%%%%%%%%%%%%%%%%%%%%%%%%%%%%%%
%%%%%%%%%%%%%%%%%%%%%%%%%%%%%%%%%%%%%%%%%%%%%%%%%%%%%%%%%%%%%%%%%%%%%%%%%%%%%%%%%%%%%%%%%%%%%%%%%%%%%%%%
%%%%%%%%%%%%%%%%%%%%%%%%%%%%%%%%%%%%%%%%%%%%%%%%%%%%%%%%%%%%%%%%%%%%%%%%%%%%%%%%%%%%%%%%%%%%%%%%%%%%%%%%
%%%%%%%%%%%%%%%%%%%%%%%%%%%%%%%%%%%%%%%%%%%%%%%%%%%%%%%%%%%%%%%%%%%%%%%%%%%%%%%%%%%%%%%%%%%%%%%%%%%%%%%%
%%%%%%%%%%%%%%%%%%%%%%%%%%%%%%%%%%%%%%%%%%%%%%%%%%%%%%%%%%%%%%%%%%%%%%%%%%%%%%%%%%%%%%%%%%%%%%%%%%%%%%%%

\section{Infinite Tensor Products and the Quantum Measurement Problem}

Infinite tensor products, when interpreted as infinite chains of nested measurements, provide a compelling framework for addressing the quantum measurement problem.
By introducing disruptions to unitary equivalence through sectorization and factorization, this approach offers a potential reconciliation between the unitary evolution of quantum systems and the apparent collapse of the wavefunction during measurement.

The quantum measurement problem remains one of the most profound challenges in quantum mechanics, arising from the apparent inconsistency between two fundamental processes identified by von Neumann in 1932~\cite{v-neumann-49,vonNeumann2018Feb,everett}. These processes are:
{
\setlength{\leftmargini}{15pt}
\begin{itemize}
    \item[] Process 1: The discontinuous, probabilistic change in a quantum state upon measurement. For a system in a superposition $\psi = \sum_{i} c_i \phi_i$, observing a quantity with eigenstates $\phi_1, \phi_2, \dots$ collapses the state to $\phi_j$ with probability $|c_j|^2$.
    \item[] Process 2: The continuous, deterministic evolution of an isolated system's state according to the Schr\"odinger equation, $\partial \psi / \partial t = U \psi$, where $U$ is a unitary operator.
\end{itemize}
}

The crux of the measurement problem is whether the unitary evolution (Process 2) can fully account for the collapse observed in measurements (Process 1), or if an additional mechanism is required. This section explores the use of infinite tensor products, interpreted as infinite nestings of Wigner's friend scenarios, as a potential resolution to this problem.

\subsection{Infinite Tensor Products in Nested Measurement Scenarios}

A promising approach to addressing the measurement problem involves infinite tensor products, which model an infinite sequence of observers, each measuring the system observed by the previous observer. This setup is reminiscent of Wigner's friend thought experiments, where the act of measurement is recursively applied. Unlike finite tensor products, infinite tensor products can disrupt unitary equivalence through mechanisms such as sectorization and factorization, potentially providing a bridge between unitary evolution and the apparent collapse of the wavefunction.

\subsection{The Von Neumann-Landau Measurement Scheme}

In the von Neumann-Landau framework, the measurement process is modeled by the interaction between an object and a measurement apparatus. The object is prepared in a state $|\psi\rangle = \sum_{i=1}^n a_i |\psi_i\rangle$, which is a superposition relative to the measurement basis. The measurement apparatus is represented by another state $|\phi\rangle = \sum_{j=1}^n b_j |\phi_j\rangle$. Upon interaction, the combined state of the object and apparatus becomes:
\[
|\Psi\rangle = \sum_{i,j=1}^n c_{ij} |\psi_i\rangle \otimes |\phi_j\rangle,
\]
where the coefficients $c_{ij}$ cannot be factorized, indicating entanglement between the object and the apparatus.

While this scheme is straightforward for finite systems, extending it to an infinite chain of measurements---where each measurement is itself measured by another observer, ad infinitum---requires the use of infinite tensor products. This extension is mathematically non-trivial and was first rigorously studied by von Neumann in 1939 \cite{vonNeumann1939}.

The construction of the infinite tensor product space proceeds as follows:
\begin{enumerate}
    \item Begin with elementary tensors of the form $\bigotimes_{n=1}^{\infty} |k_n\rangle$.
    \item Define the inner product between two elementary tensors as:
    \[
    \left\langle \bigotimes_{n=1}^{\infty} |k_n\rangle \middle| \bigotimes_{n=1}^{\infty} |l_n\rangle \right\rangle = \prod_{n=1}^{\infty} \langle k_n | l_n \rangle,
    \]
    provided the product converges; otherwise, it is zero.
    \item Consider finite linear combinations of these elementary tensors:
    \[
    \sum_i c_i \bigotimes_{n=1}^{\infty} |k_n^{(i)}\rangle,
    \]
    where $c_i$ are complex coefficients and $|k_n^{(i)}\rangle$ are basis vectors.
    \item Obtain the complete Hilbert space $\bigotimes_{n=1}^{\infty} \mathcal{H}_n$ by taking the closure of the space of finite linear combinations.
\end{enumerate}

This construction introduces several challenges that must be addressed to fully understand its implications for the measurement problem.

\subsection{Challenges with Infinite Tensor Products}

\subsubsection{Cardinality}

As pointed out by von Neumann in 1939~\cite{vonNeumann1939}, a fundamental issue with infinite tensor products is the uncountable cardinality of the resulting space.
Just as the real numbers cannot be enumerated by a countable set, the infinite tensor product space cannot be spanned by a countable basis.
Such generalizations involve nonseparable Hilbert spaces and higher set-theoretical powers of their orthonomal bases, thereby spoiling unitary equivalence with (in)finite-dimensional separable Hilbert spaces.

The interval \([0,1)\) can be represented in binary form as
\[
\left\{ 0.x_1x_2x_3\ldots \; \middle| \; x_i \in \{0,1\} \text{ for all } i \in \mathbb{N} \right\}.
\]
Here, each \(x_i\) is a binary digit (0 or 1), and the sequence extends indefinitely. This set of all infinite binary sequences is uncountable, with cardinality
\[
\#\{0,1\}^{\mathbb{N}} = 2^{\aleph_0},
\]
which, by Cantor's diagonal argument mentioned earlier, is strictly larger than the cardinality \(\aleph_0\) of the natural numbers.

In close analogy, consider an infinite sequence of qubits, where each qubit is a two-state quantum system with basis states \(\lvert 0 \rangle\) and \(\lvert 1 \rangle\). A product state in the infinite tensor product is written as
\[
\lvert x_1 x_2 x_3 \ldots \rangle = \lvert x_1 \rangle \otimes \lvert x_2 \rangle \otimes \lvert x_3 \rangle \otimes \cdots,
\]
with \(x_i \in \{0,1\}\) for all \(i \in \mathbb{N}\). The collection of all such product states corresponds exactly to the set of infinite binary sequences, hence its cardinality is also \(2^{\aleph_0}\).

A denumerable set of product states is any countable subset of the infinite tensor product states. For instance, the set
\[
\Big\{ \lvert 000\ldots \rangle,\ \lvert 100\ldots \rangle,\ \lvert 010\ldots \rangle,\ \ldots \Big\}
\]
can be put into a one-to-one correspondence with the natural numbers \(\mathbb{N}\) and thus has cardinality \(\aleph_0\). Clearly, \(\aleph_0 < 2^{\aleph_0}\).

%\subsection{Cantor's Diagonal Argument}
%Cantor's diagonal argument shows that no countable list can exhaust all infinite binary sequences. Suppose we list a countable collection of sequences:
%\[
%s_1, s_2, s_3, \ldots,
%\]
%where
%\[
%s_n = s_{n1} s_{n2} s_{n3} \ldots \quad \text{with } s_{ni} \in \{0,1\}.
%\]
%Then, construct a new sequence \(s = d_1 d_2 d_3 \ldots\) by defining
%\[
%d_i = 1 - s_{ii}.
%\]
%Since \(s\) differs from each \(s_n\) at the \(n\)th digit, it cannot appear in the list. This contradiction establishes that the set of all infinite binary sequences is uncountable.

%\subsection{Unitary Equivalence}
A Hilbert space is \emph{separable} if it has a countable (finite or infinite) orthonormal basis; otherwise it is called nonseparable.
Two Hilbert spaces $\mathcal{H}_1$ and $\mathcal{H}_2$ are unitarily equivalent if and only if they have the same \emph{dimension} (that is, their orthonormal bases share the same cardinality).
More explicitly, if \(\vert e_i \rangle\) and \(\vert f_i \rangle\) are denumerable orthonormal bases for separable \(\mathcal{H}_1\) and \(\mathcal{H}_2\), respectively, then the unitary operator is given by
\[
U = \sum_i \vert f_i \rangle \langle e_i \vert.
\]
%Similarly, for two subspaces \(V_1 \subset \mathcal{H}_1\) and \(V_2 \subset \mathcal{H}_2\), we say that \(V_1\) and \(V_2\) are unitarily equivalent if there exists a unitary operator \(U\) such that \(U(V_1) = V_2\).

In our scenario, the full Hilbert space of the infinite tensor product space is nonseparable, as it has an orthonormal basis consisting of product states with cardinality \(2^{\aleph_0}\).
Any candidate unitary operator mapping a countable (denumerable) subset of product states (with cardinality \(\aleph_0\)) to the full set must preserve the inner-product structure and be surjective onto the basis.
However, since a unitary map must preserve the cardinality of an orthonormal basis, and we have
\(
\aleph_0 < 2^{\aleph_0},
\)
no such unitary operator can exist that maps a countable subset onto the full uncountable basis.

\subsubsection{Inner Product and Orthogonality}

Another significant challenge arises in defining the inner product for infinite tensor products. For two states $|\Psi\rangle = \bigotimes_{i=1}^{\infty} | x _i\rangle$ and $|\Phi\rangle = \bigotimes_{i=1}^{\infty} | y _i\rangle$, the inner product is given by:
\[
\langle \Psi | \Phi \rangle = \prod_{i=1}^{\infty} \langle  x _i |  y _i \rangle.
\]
If each $\langle  x _i |  y _i \rangle = 1 - \epsilon_i$ with $0 < \epsilon_i \ll 1$, and if the series $\sum_{i=1}^{\infty} \epsilon_i$ diverges, then:
\[
\prod_{i=1}^{\infty} (1 - \epsilon_i) \approx \exp\left(-\sum_{i=1}^{\infty} \epsilon_i\right) \to 0.
\]
This implies that states which are only slightly different across infinitely many components can have an inner product that approaches zero, making them effectively orthogonal. Moreover, if the states differ in even a single component such that $\langle  x _k |  y _k \rangle = 0$ for some $k$, the entire inner product becomes zero, regardless of the similarity in other components. This behavior disrupts traditional notions of orthogonality and complicates the interpretation of measurement outcomes.

\subsection{Sectorization as a Solution}
\label{2025-idi-sectorization}

To address these issues, von Neumann proposed partitioning the infinite tensor product space into disjoint `regions' or \textit{sectors}---equivalence classes of states that are `close' to each other in a specific sense~\cite{vonNeumann1939}. Two states $|\Psi\rangle$ and $|\Phi\rangle$ are considered to be in the same sector if:
\[
\sum_{i=1}^{\infty} (1 - |\langle  x _i |  y _i \rangle|) < \infty.
\]
This condition ensures that the states differ significantly in only finitely many components. These sectors can be thought of as corresponding to distinct macroscopic or classical outcomes, potentially offering a way to interpret measurement results within the framework of unitary evolution~\cite{Grangier-2020,van-den-bossche-2023-a,van-den-bossche-2023-b,van-den-bossche-2023-c}.

\subsection{Factorization and Unitary Equivalence}

A further opportunity arises from the entanglement of infinite components, which can lead to different types of factors (e.g., type I, II, or III in von Neumann algebra classification) that are not unitarily equivalent. This lack of unitary equivalence suggests a mechanism by which the infinite tensor product space can accommodate irreversible processes, such as those observed in quantum measurements.

\subsection{Role of equivalence classes}

In sectorization, equivalence classes are employed to partition the infinite tensor product space into distinct sectors, each comprising states that are equivalent modulo differences in only finitely many components.
This classification is pivotal for associating each sector with a specific, classical measurement outcome, thereby offering a framework to reconcile the continuous, unitary evolution of quantum systems with the discrete nature of observed measurement results.
Furthermore, the lack of unitary equivalence between different sectors---stemming from the factorization of the space---underscores the critical role of equivalence classes in establishing the irreversibility characteristic of the quantum measurement process.
By defining these equivalence classes, sectorization simplifies the handling of complex quantum systems and provides insight into the transition from quantum superpositions to definite classical states.

%%%%%%%%%%%%%%%%%%%%%%%%%%%%%%%%%%%%%%%%%%%%%%%%%%%%%%%%%%%%%%%%%%%%%%%%%%%%%%%%%%

\section{Infinite Precision Microstates and the Emergence of Macroscopic Irreversibility}

A common starting point in statistical physics is to describe an isolated many-particle system by specifying its \emph{microstate} with infinite precision.
In principle, if every particle's position and momentum were known exactly, the time-reversible microscopic laws (that is, Newtonian or unitary quantum dynamics) imply that every evolution has a time-reversed twin.
This observation is at the heart of Loschmidt's \textit{Umkehreinwand} (reversal objection)~\cite{darrigol-2021}:
If one were able to precisely reverse the velocities of all particles, then every macroscopic process (such as the free expansion of a gas) would be exactly reversible.
In other words, entropy would remain constant with infinite precision.

In an extreme scenario where a hidden entity, such as Maxwell's demon, manipulates a system at the microphysical level with infinite precision---unbeknownst
to observers who are limited to finite precision macroscopic measurements---the demon could orchestrate processes like the spontaneous unmixing of two previously mixed gases.
This would make entropy appear to decrease from the observers' macroscopic perspective, creating the illusion of a contradiction with the second law of thermodynamics,
which states that entropy in an isolated system cannot decrease over time.

However, concepts such as physical means and demons manipulating microstates with infinite precision is an idealized notion.
In reality, our ability to specify, measure, or manipulate microscopic degrees of freedom in any physical system is operationally limited.
These practical constraints mean that microstates can only be defined with \emph{finite} precision.
Consequently, when attempting to reverse a system's evolution, the unavoidable small uncertainties are amplified through the system's complex (often chaotic) dynamics.

Moreover, the concept of \emph{means-relative reversibility}  emphasizes that while the microscopic laws are symmetric, the notion of a reversible process depends on the precision and the scale at which the state is defined.
Maxwell's pragmatic approach---``avoiding all personal inquiries [[about individual molecules]] which would only get me into trouble''~\cite{Maxwell-1879,garber}---illustrates
that the coarse-grained description relevant for thermodynamics deliberately sidesteps the need for infinite precision. In this framework, macroscopic irreversibility emerges from the overwhelming statistical likelihood that a system will evolve toward states of higher entropy,
even though the underlying equations are time-symmetric.

%Indeed, as an elementary probabilistic urn argument shows, microphysically  entropy might even decrease~\cite{Ehrenfest_06g}:
%take, for example, two urns with balls in them, in a microstate that has most balls in the first urn, and a very small number of balls in the second urn.
%Now if there is some transition probability that per timestep any ball of one urn makes it into the other urn (which is constant for all balls),
%and we let the system evolve, then most likely we will find an almost 50:50 ratio of balls in both urns.
%But even in that `relaxed' most likely state it is possible and even necessary that after an `enormously long time' all balls will have assembled in a single one urn.
%So, if one continues evolution long enough, it would ultimately be downright enormously improbable if these high and highest `outliars' [\textit{Buckel}~\cite{Ehrenfest_06g}] were to remain absent.
%But that suffices for Zermelo's \textit{Wiederkehreinwand} (recurrence objection), based on Poincar\'e's recurrence theorem.
%If the system is capable of universal computation we might argue that recurrence times for certain states---such as the ones associated with
%the halting probability Omega discussed earlier---`grows beyond any specifiable size'; indeed, faster than any recursive (computable) lower bound~\cite{svozil-93}.

\begin{figure*}[htbp] % Placement suggestion: here, top, bottom, page
    \centering % Center the figure horizontally
    \includegraphics[width=0.9\textwidth]{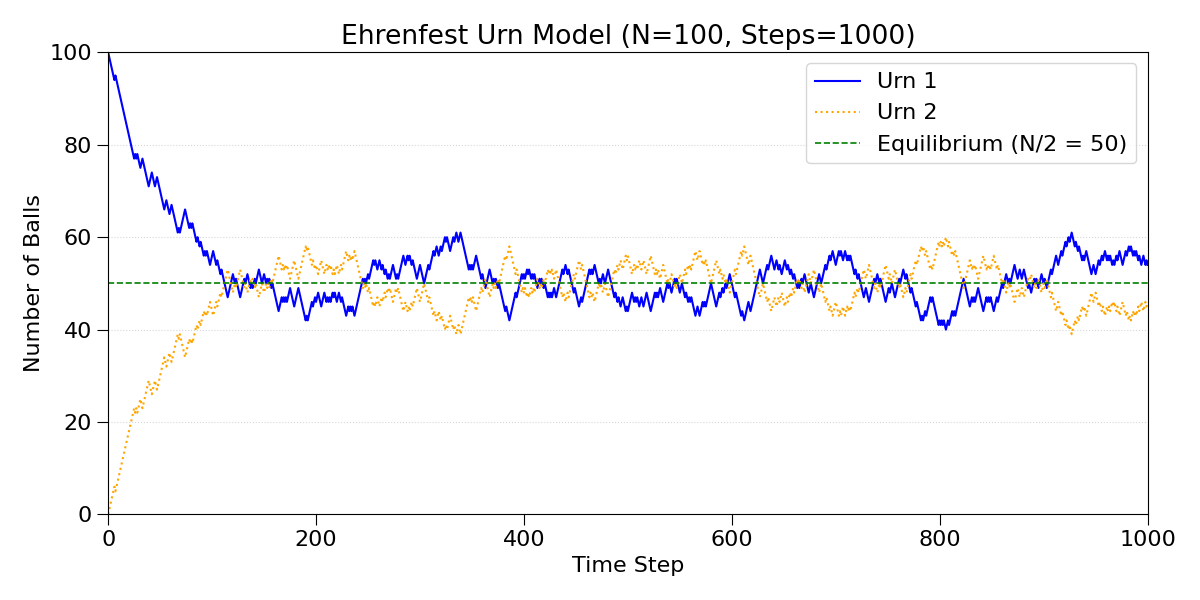}
    % Include the graphic file. Adjust 'width' as needed (e.g., \textwidth, 10cm).
    % Make sure the image file "2025-idi-Figure_1.png" is in the same directory
    % as your .tex file, or provide a relative/absolute path.
    \caption{Evolution of the number of balls in Urn 1 (blue) and Urn 2 (dotted, orange) in the Ehrenfest Urn Model (N=100, 1000 steps), starting from a low-entropy state with Urn 1 filled and Urn 2 empty.
The dashed line marks the equilibrium state (N/2=50). The simulation highlights the system's relaxation towards equilibrium and the persistent fluctuations around it, illustrating the microscopic reversibility that underlies Zermelo's recurrence objection.}
%Simulation of the Ehrenfest Urn Model with N=100 balls over 1000 time steps. The system starts with all balls in Urn 1 (n1=100). The plot shows the number of balls in Urn 1 (n1) as a function of time steps. The dashed line indicates the equilibrium state (n1=N/2=50). The simulation illustrates the rapid approach towards equilibrium followed by statistical fluctuations around it, consistent with the microscopic reversibility underlying Zermelo's recurrence objection.}
    \label{fig:ehrenfest_sim} % Label for cross-referencing using \ref{fig:ehrenfest_sim} or \Cref{fig:ehrenfest_sim} (with cleveref package)
\end{figure*}

The Ehrenfest urn model~\cite{Ehrenfest_06g} provides an elementary probabilistic illustration suggesting that entropy, viewed microphysically, might even decrease.
Consider two urns initially containing an uneven distribution of balls, with most in the first urn.
Assume a constant probability per time step for any ball to transfer to the other urn.
As the system evolves, it will most likely approach an equilibrium state with roughly equal numbers of balls (a 50:50 ratio) in both urns, corresponding to maximum entropy in this analogy.
However, even from this high-probability equilibrium state, fluctuations are possible.
Poincar\'e's recurrence theorem implies that after an `enormously long time,' it is not only possible but inevitable that the system will return to highly improbable states, such as having all balls collected in a single urn.
Therefore, the eventual reappearance of these low-entropy configurations (`outliers' or \textit{Buckel}\cite{Ehrenfest_06g}) cannot be ruled out; indeed, their absence over sufficiently long timescales would be extremely improbable.
This guaranteed recurrence forms the basis of Zermelo's \textit{Wiederkehreinwand} (recurrence objection) against monotonic entropy increase at maximal (microphysical) resolution.
The simulation results depicted in Figure~\ref{fig:ehrenfest_sim} demonstrate the system's tendency towards a state of higher entropy (equilibrium), punctuated by fluctuations that manifest as temporary, occasional decreases in entropy.
Moreover, if the system is capable of universal computation,
recurrence times for certain `computationally complex, resource-intensive' states---such as those associated with the halting probability Omega mentioned earlier---can be expected to grow `beyond any specifiable size',
potentially faster than any recursive (computable) lower bound~\cite{svozil-93}.

In statistical mechanics, a macroscopic state is essentially an equivalence class, where the equivalence relation is defined by the condition that two microscopic states are considered equivalent if they share the same values for macroscopic variables, such as energy or volume.
In the aforementioned example, macroscopic states with roughly equal numbers of balls (a 50:50 ratio) in both urns are much more likely than macroscopic states characterized by outliers with the same number of balls.
Therefore, macroscopic systems tend to evolve toward entropy increase.

For example, in a gas, all possible molecular arrangements that result in the same pressure and temperature belong to the same macroscopic state.
Consequently, microscopic states can be formally `bundled together' or `grouped' based on macroscopic equivalence:
If they cannot be distinguished through operational means at the macroscopic level, they are defined as equivalent.
The corresponding binary equivalence relation, applied to microstates, naturally satisfies the properties of reflexivity, symmetry, and transitivity.

In summary, while the mathematical description of a system in terms of infinite precision microstates leads to reversible trajectories, the physical impossibility of achieving such precision guarantees that real systems display irreversible behavior.
The irreversible macroscopic laws of thermodynamics are thus understood as emergent, effective descriptions that arise from practical limitations on precision and the statistical averaging over an enormous number of microstates.
The physical means define an equivalence relation on microphysical states. The corresponding equivalence classes can be identified with macroscopic states.

\section{Formalization of FAPPness by equivalence relations}

In an early critique of sectorization-type arguments~\cite{hepp-1972,bub-2015} reviewed in Section~\ref{2025-idi-sectorization}, Bell argued~\cite{Bell-1975} that unlimited or even actually infinite means are physically unattainable.
He later introduced the related concept of For-All-Practical-Purposes (FAPP)~\cite{bell-a},
which replaces transfinite means with finite, physically operational ones.

The concept of FAPP indistinguishability can be rigorously formalized using equivalence relations. In different physical contexts, these equivalence relations partition microscopic configurations into equivalence classes, grouping together states that are operationally indistinguishable at a higher level of description. The three primary instantiations of such equivalence classes, as discussed in previous sections, are:

\subsection{Classical Analysis}
In classical mechanics and dynamical systems, coarse-graining leads to an effective partitioning of phase space into equivalence classes. Two microstates belong to the same class if they yield identical macroscopic observables within a given resolution limit. This follows naturally from measurement constraints and computational limitations in practical analysis.

\subsection{Sectorization in Quantum Mechanics}
In quantum theory, the emergence of classical-like behavior is often described using \textit{superselection sectors} or \textit{decoherence-induced equivalence classes}. Here, quantum states that differ only by superpositions within a decohered basis (due to environmental interactions) become practically indistinguishable. Such states effectively belong to the same equivalence class, as they do not interfere and cannot be resolved through macroscopic measurements.

\subsection{Macrostates in Statistical Physics}
In statistical mechanics, a macroscopic state corresponds to an equivalence class of microstates that share the same macroscopic variables, such as energy, volume, or magnetization. Since individual microstates fluctuate rapidly and are inaccessible in practice, all configurations that yield the same macroscopic properties are grouped together, forming a thermodynamic macrostate.

In all three cases, the corresponding equivalence relation on microstates satisfies reflexivity, symmetry, and transitivity, ensuring a well-defined partitioning of state space. This formalization captures the essence of FAPP reasoning, where practical indistinguishability justifies the use of equivalence classes in physical descriptions.

\section{Conclusion}

This paper explored the role of infinity in bridging microscopic and macroscopic descriptions in physics, focusing on the emergence of irreversibility from reversible dynamics. We examined how infinite processes are essential in mathematical constructions, such as the transition from rational to real numbers, and how they manifest in physical theories, from statistical mechanics to quantum measurement.

In classical analysis, infinite precision is a theoretical idealization that is unattainable in practice. The necessity of coarse-graining and finite resolution in measurements leads naturally to the formation of equivalence classes that group together states indistinguishable for all practical purposes (FAPP). This provides a foundation for understanding macroscopic irreversibility despite the underlying time-reversible microscopic laws.

In quantum mechanics, infinite tensor products and sectorization offer a framework for understanding the transition from unitary evolution to apparent wavefunction collapse.
Von Neumann's insights~\cite{vonNeumann1939} emphasize that the full set of product states in an infinite tensor product is uncountably infinite, with a cardinality of \(2^{\aleph_0}\).
The resulting space is nonseparable.
This sharply contrasts with any countable subset, which has a cardinality of \(\aleph_0\); here, the distinction reflects the difference between nonseparability and separability.
Since unitary operators preserve the inner product structure---and, consequently, the cardinality of any orthonormal basis---no unitary transformation can map a countable subset onto the full uncountable set.
Therefore, under constraints such as (finite or infinite) denumerable  group actions, unitary equivalence fails in the limit of infinite tensor products.
The partitioning of Hilbert space into equivalence classes through decoherence-induced superselection rules highlights how quantum-to-classical transitions can emerge from infinite degrees of freedom.

In statistical mechanics, macrostates are equivalence classes of microstates that share the same macroscopic observables, such as energy or volume. The practical impossibility of resolving individual microstates supports the statistical interpretation of thermodynamic irreversibility.

By formalizing Bell's FAPP approach using equivalence relations, we provided a unifying perspective on how operational indistinguishability underlies emergent macroscopic behavior. Across classical analysis, quantum mechanics, and statistical physics, equivalence classes play a crucial role in describing physical reality at different scales. This perspective underscores the foundational role of infinity in physics and its implications for the nature of measurement, irreversibility, and emergent phenomena.

%%%%%%%%%%%%%%%%%%%%%%%%%%%%%%%%%%%%%%%%%%%%%%%%%%%%%%%%%%%%%%%%%%%%%%%%%%%%%%%%%%%%%%%%%%%%%%%%%%%%%%%%%%%%%%%%%%%%%%%%%%%%%%%%%%%%
\ifws
%%%%%%%%%%%%%%%%%%%%%%%%%%%%%%%%%%%%%%%%%%
\vspace{6pt}

\funding{
This research was funded in whole or in part by the Austrian Science Fund (FWF) [Grant DOI:10.55776/I4579].
%The author acknowledges TU Wien Bibliothek for financial support through its Open Access Funding Programme.
}

\acknowledgments{
%The author gratefully acknowledges discussions with Mohammad Hadi Shekarriz.

%Josef Tkadlec has kindly provided a program to find all two-valued states and important properties thereof, such as (non)separability,given the set of contexts of a logic.

}

\conflictsofinterest{The author declares no conflicts of interest.
The funders had no role in the design of the study; in the collection, analyses, or interpretation of data;
in the writing of the manuscript; or in the decision to publish the results.}

%%%%%%%%%%%%%%%%%%%%%%%%%%%%%%%%%%%%%%%%%%
%% Optional

%% Only for journal Encyclopedia
%\entrylink{The Link to this entry published on the encyclopedia platform.}

%%%%%%%%%%%%%\abbreviations{Abbreviations}{
%%%%%%%%%%%%%The following abbreviations are used in this manuscript:\\
%%%%%%%%%%%%%
%%%%%%%%%%%%%\noindent
%%%%%%%%%%%%%\begin{tabular}{@{}ll}
%%%%%%%%%%%%%MDPI & Multidisciplinary Digital Publishing Institute\\
%%%%%%%%%%%%%DOAJ & Directory of open access journals\\
%%%%%%%%%%%%%TLA & Three letter acronym\\
%%%%%%%%%%%%%LD & Linear dichroism
%%%%%%%%%%%%%\end{tabular}
%%%%%%%%%%%%%}

%%%%%%%%%%%%%%%%%%%%%%%%%%%%%%%%%%%%%%%%%%
%\isPreprints{}{% This command is only used for `preprints'.
\begin{adjustwidth}{-\extralength}{0cm}
%} % If the paper is `preprints', please uncomment this parenthesis.

\reftitle{References}


 \bibliography{svozil}

\PublishersNote{}
%\isPreprints{}{% This command is only used for `preprints'.
\end{adjustwidth}
%} % If the paper is `preprints', please uncomment this parenthesis.

%%%%%%%%%%%%%%%%%%%%%%%%%%%%%%%%%%%%%%%%%%%%%%%%%%%%%%%%%%%%%%%%%%%%%%%%%%%%%%%%%%%%%%%%%%%%%%%%%%%%%%%%%%%%%%%%%%%%%%%%%%%

\else

\begin{acknowledgments}
The author gratefully acknowledges discussions with Noson S. Yanofsky.
I am grateful to an anonymous referee for drawing my attention to surreal numbers in this context.
%
%Josef Tkadlec has kindly provided a program to find all two-valued states and important properties thereof, such as (non)separability,
%given the set of contexts of a logic.
%
This research was funded in whole or in part by the Austrian Science Fund (FWF) [Grant DOI:10.55776/I4579].
%The author acknowledges TU Wien Bibliothek for financial support through its Open Access Funding Programme.
\end{acknowledgments}

\bibliography{svozil}

%%%%%%%%%%%%%%%%%%%%%%%%%%%%%%%%%%%%%%%%%%%%%%%%%%%%%%%%%%%%%%%%%%%%%%%%%%%%%%%%%%%%%%%%%%%%%%%%%%%%%%%%%%%%%%%%%%%%%%%%
\fi

\end{document}